\documentclass[a4paper,fleqn]{cas-dc}

\usepackage[authoryear]{natbib}
\usepackage{lipsum}

\usepackage[switch]{lineno}

\def\tsc#1{\csdef{#1}{\textsc{\lowercase{#1}}\xspace}}
\tsc{WGM}
\tsc{QE}

\begin{document}
\let\WriteBookmarks\relax
\def\floatpagepagefraction{1}
\def\textpagefraction{.001}

\shorttitle{SWOT Water Detection using Dynamic GNNs}    

\shortauthors{Baumann et al.}  

\title [mode = title]{Toward Enhanced Water Detection in SWOT Pixel Clouds using Dynamic Graph Neural Networks}  



%

\author[1,2]{Christoph Baumann}[orcid=0009-0000-7433-9857]
\cormark[1]

\ead{baumanch@ethz.ch}

\credit{Conceptualization, Methodology, Software, Validation, Formal analysis, Investigation, Data curation, Visualization, Writing - original draft, Writing - review \& editing}

\affiliation[1]{organization={Institute of Geodesy and Photogrammetry, ETH Zurich},
            addressline={Robert-Gnehm-Weg 15}, 
            city={Zurich},
            postcode={8093},
            country={Switzerland}}

\author[1]{Junyang Gou}[orcid=0000-0002-7599-0577]
\cormark[1]

\ead{jungou@ethz.ch}

\credit{Conceptualization, Methodology, Validation, Investigation, Supervision, Writing - review \& editing}

\affiliation[2]{organization={Deutsches Geodätisches Forschungsinstitut, Technische Universität München (DGFI-TUM)},
            addressline={Arcisstr. 21}, 
            city={München},
            postcode={80333},
            country={Germany}}

\author[2]{Christian Schwatke}[orcid=0000-0002-4741-3449]

\credit{Validation, Supervision, Writing - review \& editing}

\author[3]{Mohammad J. Tourian}[orcid=0000-0002-4200-0848]

\credit{Conceptualization, Supervision, Writing - review \& editing}

\affiliation[3]{organization={Institute of Geodesy, University of Stuttgart},
            addressline={Geschwister-Scholl-Str. 24D}, 
            city={Stuttgart},
            postcode={70174}, 
            state={},
            country={Germany}}
            
\author[2]{Florian Seitz}[orcid=0000-0002-0718-6069]

\credit{Supervision, Writing - review \& editing}

\author[1]{Benedikt Soja}[orcid=0000-0002-7010-2147]

\credit{Conceptualization, Supervision, Writing - review \& editing}

\cortext[1]{Corresponding author}



\begin{abstract}
The Surface Water and Ocean Topography (SWOT) mission offers unprecedented freshwater monitoring capabilities through its innovative wide-swath measurement system, which generates several data products, including the high-resolution pixel cloud (PIXC) product. However, the native PIXC water classification remains prone to systematic misclassification in urban environments, where strong radar returns from non-water surfaces are the predominant error sources. We present a deep learning approach that enhances land-water classification directly on SWOT PIXC data based on a dynamic graph convolutional neural network that simultaneously exploits spatial proximity and feature similarity to derive binary land-water class labels. The model is trained on a full year of PIXC data for the Dallas-Fort Worth metropolitan area using pixel-level ground-truth class labels derived from the DSWx-HLS product, which provides water-class labels based on Landsat and Sentinel-2 at a 30\,m spatial resolution. Against the DSWx-derived reference, the proposed method increases the mean scene-level F1 score against the DSWx-derived reference from 0.52 to 0.86 on the temporally independent test set and from 0.43 to 0.74 on the spatiotemporal test set, relative to the native PIXC classification. These results demonstrate that dynamic graph neural networks are well-suited to the irregular, point-cloud-like structure of PIXC data and offer a scalable path toward more reliable urban flood monitoring and freshwater mapping at the high spatial resolution provided by the SWOT PIXC product.
\end{abstract}




\maketitle

\section{Introduction}\label{sec:intro}

Water represents an indispensable need for life on Earth \citep{cretaux_lake_2016}. Consequently, monitoring, safeguarding, and managing water resources become tasks of paramount importance. This takes many forms, ranging from the protection of ecosystems and biodiversity associated with freshwater resources \citep{dudgeon_freshwater_2006, wolfram_water_2021, chapman_role_2022}, to safeguarding human health and water use \citep{sagan_monitoring_2020, ngamile_trends_2025}, security, and drought and flood management~\citep{Gou2024GlobalHighResolution}. Changes in water storage can have far-reaching implications for drinking water availability, irrigation, and power supplies through hydropower \citep{donchyts_high-resolution_2022}. 

Satellite observations enable monitoring at scale, irrespective of physical access or in situ infrastructure. First efforts in monitoring water resources using satellite technology date back to the 1970s and 1980s \citep{musa_review_2015}. By the 1990s, the number of sensors and datasets had increased significantly \citep{lettenmaier_inroads_2015}. Since the late 1990s, radar altimetry, in conjunction with optical and microwave imaging, has enabled observations of water level, surface water extent, water storage, and water quality \citep{sheffield_satellite_2018}. With the rapid expansion in capabilities and scientific use over the last two to three decades, they are today considered vital components of the monitoring system for the global water cycle and water security assessment~\citep{tourian2022hydrosat}.

Well-established methods are found in the domains of multi-spectral satellite imagery \citep{McFeeters1996TheFeatures, Feyisa2014AutomatedImagery, Xu2006ModificationImagery, Elmi2016RS, Pekel2016High-resolutionChanges} and Synthetic Aperture Radar (SAR) imagery \citep{Liang2019A, Nemni2020Fully, li_u-net-based_2023, Amitrano2024Flood} for both water surface detection and, more recently, flood monitoring, including several studies successfully establishing deep learning-based detection approaches \citep{Guo2022Water-Body}. While multi-spectral imagery offers rich spectral information with high spatial resolution, it remains susceptible to cloud cover, limiting its observational capabilities for continuous monitoring \citep{elmi2023retrieving}. SAR overcomes this limitation with its all-weather capability, yet is fundamentally constrained to measuring surface extent rather than providing physical water level information. Traditional satellite altimetry addresses this gap by directly measuring water surface elevation, though its limitation lies in the inability to directly observe surface extents. Furthermore, the quality of altimetry measurements degrades sharply in nearshore areas due to contamination in the returned pulses, which further complicates accurate detection of water bodies \citep{Gou2022RiwisarSwhA,ke2025novel,Tourian-2025,khalili20264d}.

The launch of the Surface Water and Ocean Topography (SWOT) mission represents a milestone in the evolution of satellite altimetry observations. As the first wide-swath altimeter, SWOT combines swath-based coverage with cloud-cover resistance and physically meaningful height measurements, enabling unprecedented monitoring capabilities of inland water bodies \citep{Biancamaria2016TheHydrology, Fu2024TheWater}. This has already enabled remarkable research achievements, such as near-global shape and storage assessments of rivers at unprecedented spatial scales \citep{cerbelaud_wide-swath_2026}, the quantification of river tidal dynamics \citep{hart-davis_observing_2026}, and demonstrated improvements in volume monitoring for small lakes \citep{jing_exploring_2026}. As SWOT continues to demonstrate advances at finer spatial scales, extracting reliable information at the highest possible spatial resolution presents an essential task.

The classification algorithm natively employed in the SWOT Level-2 High-Rate Pixel Cloud (PIXC) product is based on maximum a posteriori (MAP) probability estimation. Markov Random Fields are leveraged to enforce spatial regularity, and water detection is performed iteratively based on estimated coherent power \citep{Lobry2019WaterDetection, JPL-D-105504-2023}. First studies validating classification performance on post-launch data identify incidence angle sensitivity, soil moisture, and wind as primary drivers of reduced accuracy \citep{BONASSIES2026115101, Gasnier2024}. A dominant source of omission errors is the \textit{dark water} phenomenon \citep{Li2026, JPLD-1095322024SWOTHandbook}, a compounded effect of incidence angle sensitivity and low surface roughness under calm wind conditions. At the near-nadir incidence angles of the Ka-band radar interferometer (KaRIn), the primary instrument on board the SWOT satellite, a calm surface acts specularly, reflecting the radar signal away from the sensor, producing a backscatter return substantially lower than the typical 10\,dB water-land contrast \citep{fjortoft_karin_2014}. Conversely, commission errors arise when non-water surfaces produce high-powered returns, commonly referred to as \textit{bright land} \citep{JPLD-1095322024SWOTHandbook}. Daily SWOT Cal/Val observations showed that irrigation increased backscatter by an average of 4.3\,dB on the day of irrigation \citep{Bazzi2026Observing}. Similarly, urban infrastructure is known to be an additional source of elevated radar returns. Consequently, complex surface conditions remain a persistent challenge and give rise to the need for improved classification methods \citep{BONASSIES2026115101}.

Several studies have sought to address these shortcomings. \citet{Gasnier2024} identified pathways for algorithmic development of the operational water detection algorithm, though the classifier remains coherence-based without external reference supervision. At the raster product level, \citet{Li2026} proposed an optical--radar fusion framework that compensates for dark water through backscatter enhancement and addresses water--wetland confusion via multiscale image decomposition, demonstrating improved separability validated against Sentinel-2 imagery. Similarly, \citet{Xu2026SDNet} developed SDNet, a Transformer-based framework that suppresses stripe noise via frequency-domain filtering and recovers water pixels lost to quality control flag failures. However, both approaches operate on gridded raster representations of SWOT data, and neither employs an operationally produced reference product as training supervision. Addressing classification errors at the PIXC level is further complicated by the spatially irregular nature of the point cloud. Unlike the regular gridded rasters of multispectral or SAR imagery, where convolutional neural network (CNN) architectures \citep{Lecun1998Gradient-basedRecognition} have proven highly effective for water body extraction \citep{Guo2022Water-Body}, convolutional structures are ill-suited for unstructured PIXC data without rasterization, which introduces interpolation artifacts and discards native per-pixel physical observables.

In this study, we investigate an approach to water detection directly in SWOT pixel cloud data that combines spatial similarity with feature similarity derived from physical measurement quantities to address the aforementioned limitations to classification performance at the native PIXC-level, which, to our knowledge, has not been attempted before in the literature. We aim to achieve this through a two-step graph neural network (GNN) approach that first considers spatial proximity through attention-based learning \cite{Velickovic2017GraphNetworks}, then dynamically adapts the graph structure based on similarity in the feature space, and finally performs an edge-convolutional step \citep{Wang2019DynamicClouds}. We refer to this combined mechanism as dynamic attention graph convolution (DAGC). This combination aims to reduce misclassifications due to high-powered land returns while simultaneously respecting sharp class boundaries at shorelines through the feature similarity mechanism, and further examines both the spatial and temporal generalizability of our model.

\section{Data} \label{sec:data}
\begin{figure}
    \centering
    \includegraphics[width=\linewidth]{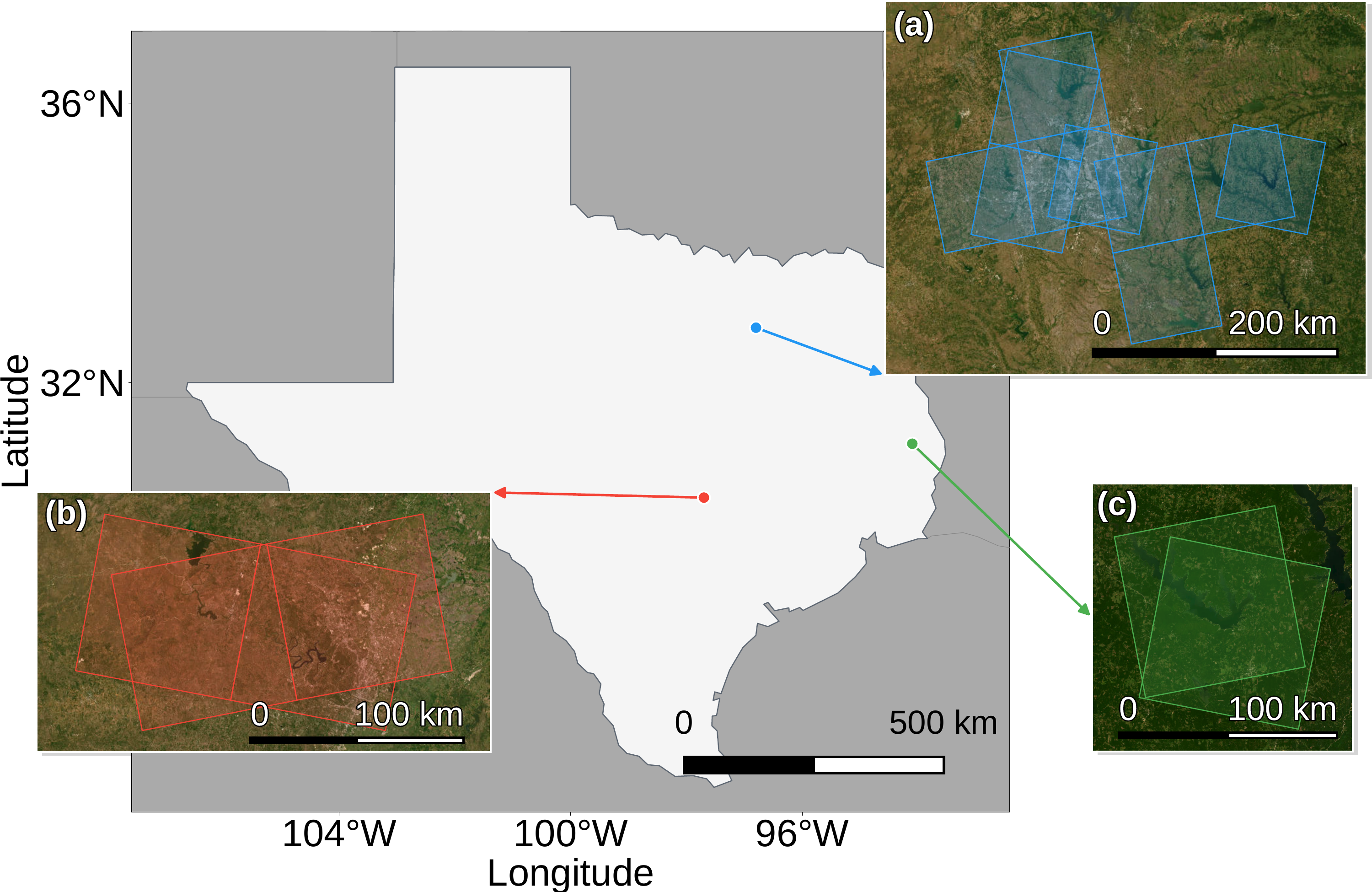}
    \caption{Geographical focus of the study with tiles used for training, validation, and testing: (a) DFW tiles: 009\_212R, 009\_213R, 022\_096L, 022\_097L, 022\_097R, 287\_212L, 300\_097L, 300\_097R, 300\_098R, 565\_212L (b) Austin tiles: 009\_208L, 009\_208R, 328\_101L, 328\_101R (c) Sam Rayburn Reservoir tiles: 259\_209R, 578\_100L. Basemap: OpenStreetMap contributors, CARTO; inset imagery: Esri; SWOT tiles: SWOT Science Team (2024).}
    \label{fig:dallas_tiles}
\end{figure}

The geographical focus of this study lies on the Dallas-Fort Worth (DFW) and Austin metropolitan areas in Texas (USA), alongside a smaller subset near Sam Rayburn Reservoir, located in eastern Texas. The geographical context is visualized in Figure \ref{fig:dallas_tiles}. All data were acquired in the years 2024 and 2025. This choice of study region provides a mix of highly urbanized environments, reservoirs of various sizes, some agricultural fields as part of the immediate surroundings of the metropolitan areas, and heavily forested areas as part of the Sam Rayburn subset. Ultimately, this enables a sufficient number of water pixels to facilitate efficient model training while incorporating the core challenge of high-powered returns from built surfaces. To obtain reliable training labels, we leverage established multispectral water detection products, which are well-suited for this purpose due to their proven performance and high spatial resolution, but are inherently limited by cloud cover. While SWOT PIXC, by contrast, penetrates cloud cover and directly measures water surface elevation, which may be beneficial in a multitude of applications, its native classification is unreliable under the conditions specifically targeted in this study. This interplay makes multispectral-derived labels a natural source of supervision.

\subsection{SWOT Pixel Clouds}\label{sec:pixc_data}
The Level-2 High-Rate PIXC product \citep{swot_swot_2025}, also referred to as L2\_HR\_PIXC or the PIXC product, is a SWOT data product specifically optimized for hydrological studies. In this study, we employ product version D. The measurement is obtained using the innovative Ka-band radar interferometer (KaRIn). Hence, it enables the acquisition of ground surface data under all meteorological conditions. It should be noted, however, that the Ka-band radar frequency, employed by the KaRIn instrument, is higher than that of the Ku-band, which is more commonly utilized in radar altimetry. It is anticipated that performance may be influenced by phenomena such as heavy rain \citep{fjortoft_karin_2014}. 

The PIXC product provides high-rate measurements with a typical pixel ground spacing of 10\,m to 70\,m across-track and 20\,m along-track, alongside a multitude of per-pixel attributes. Individual scenes cover an area of 64\,km by 64\,km.  The relevant attributes to this study are coherent power, phase noise standard deviation, number of effective medium looks, and height. Coherent power offers several advantages over other power measures, such as the incoherent average of the two power channels or the individual channel powers, including a signal power boost relative to the thermal noise floor \citep{JPL-D-105504-2023}. The phase noise standard deviation $\sigma_\phi(x)$, for each pixel $x$ quantifies the uncertainty in the interferometric phase measurement: 
\begin{equation}
    \sigma_\phi(x)=\sqrt{ \frac{\left(1-\gamma(x)^2\right)}{L_{\text{med}}(x) \cdot 2 \gamma(x)^2}},
\end{equation}
where $\gamma$ is the coherence and $L_{\text{med}}$ the number of medium looks \citep{Rosen2000SyntheticInterferometry, JPLD-1095322024SWOTHandbook}. The number of medium looks arises from the PIXC processing chain. The height measurement is derived using KaRIn interferometry to quantify surface height above the ellipsoid \citep{Peral2024}. In addition to the SWOT-based attributes, we also include prior water occurrence information, which is provided in the PIXC product. The information is based on \citet{Pekel2016High-resolutionChanges} and grounded in 32 years of Landsat imagery at approximately a 30\,m spatial resolution.

In addition to the features of the PIXC product, geolocation information from the L2\_HR\_PIXCVec product is employed \citep{swot_pixcvec_2025}. This provides us with the benefit of less noisy, height-constrained geolocation information compared to the native PIXC data. The PIXCVec product represents a one-to-one mapping with the PIXC product. Furthermore, it incorporates additional prior information by mapping water pixels back to the Prior River Database (PRD/SWORD, \cite{sword}) or the Prior Lake Database (PLD, \cite{pld}) \citep{JPLD-1095322024SWOTHandbook}.

\subsection{OPERA DSWx-HLS Product}\label{sec:opera}
The Observational Products for End-Users from Remote Sensing Analysis (OPERA) project was launched in 2021 by NASA's Jet Propulsion Laboratory (JPL). It provides a broad range of product suites covering multiple domains. In this study, we rely on the Dynamic Surface Water Extent (DSWx) product as the source of ground truth for supervised learning. The product is based on harmonized Landsat-8 and Sentinel-2A/B (DSWx-HLS) observations \citep{opera_opera_2023}.

The DSWx-HLS product employs a multistage workflow to translate multispectral surface reflectance measurements into discrete water classes. This involves computing multiple spectral indices focused on detecting water, as well as partial water in a pixel, employing auxiliary data to mitigate false-positive detections primarily caused by terrain-induced shadows or low-reflectance vegetation, and aerosol and cloud masking \citep{jones_improved_2019}. The final product comes in 30\,m by 30\,m raster format. Water class information is provided in two layers: a multi-class classification layer, which differentiates between water and partial water surfaces, and a binary water label layer. Each pixel is additionally accompanied by a confidence score given in a separate layer. In this study, we use the binary water label and confidence layers, designated \textit{BWTR} and \textit{CONF}, respectively, in the data product.

\section{Methodology}
\subsection{Dynamic Graph Neural Network for Water Body Detection} \label{sec:themodel}

The proposed binary classification model leverages a GNN to exploit spatial correlations inherent in PIXC data. Three key properties motivate this choice: first, neighboring pixels are likely to share the same class label, making spatial context informative for classification; second, PIXC observations do not reside on a regular raster grid, as pixel spacing varies across scenes; third, graph representations naturally accommodate both node-level features (per-pixel observables) and edge-level features (pairwise geometric or radiometric relationships), offering flexibility that regular convolutions cannot provide.

We represent each PIXC scene as a neighborhood graph, connecting each pixel to its ten nearest neighbors in geographic space. This construction is deliberately local, restricting the receptive field of the initial aggregation step to the most immediate spatial context. We do not assume that nearby pixels necessarily share the same class; rather, the graph structure defines which pixels exchange information, while the model learns from their feature contrasts. This is especially relevant at land-water boundaries, where adjacent pixels can exhibit markedly different characteristics despite their proximity.

\begin{figure*}
	\centering
	\includegraphics[width=.9\textwidth]{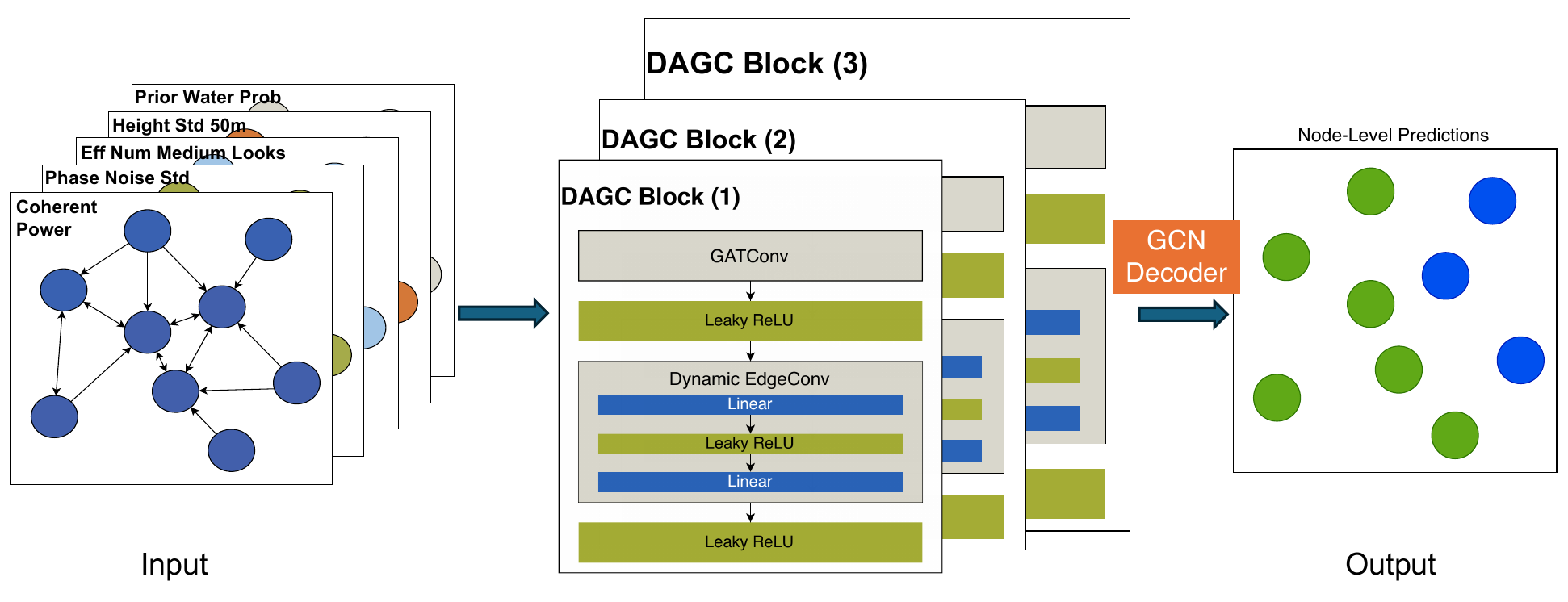}
	\caption{Schematic visualization of the foundational DAGC-based architecture, employing a set of input features, followed by a sequence of three DAGC blocks of increasing hidden layer sizes (64-128-256), and a GCN decoder to produce node-level label predictions.}
	\label{FIG:model}
\end{figure*}

GNNs operate on data represented as a graph $\mathcal{G}=(\mathcal{V}, \mathcal{E})$, with $\mathcal{V}$ denoting the set of nodes and $\mathcal{E}$ denoting the set of edges. Each node $i \in \mathcal{V}$ carries a feature vector $\mathbf{h}_i$. Learning proceeds through iterative message passing \citep{Gilmer2017NeuralChemistry}, where each node aggregates information from its local neighborhood $\mathcal{N}(v_i)$ to update its representation. The form of aggregation defines the GNN variant. Depending on the problem statement, the choice of how messages are aggregated may differ. In this study, we address this through graph attention convolution (GATConv;~\citealt{Velickovic2017GraphNetworks}), which learns normalized attention coefficients $\alpha_{ij}$ to weigh each neighbor's contribution.

In GATConv, the features of each node $i$, denoted $\mathbf{h}_i \in \mathbb{R}^F$, where $F$ is the number of features, are linearly transformed as:
\begin{equation}
    \tilde{\mathbf{h}}_i=\mathbf{W h}_i,
\end{equation}
where $\mathbf{W} \in \mathbb{R}^{F' \times F}$ is a learnable weight matrix shared across all nodes. The attention mechanism is realized as a single feedforward network layer with weight vector $\mathbf{a} \in \mathbb{R}^{2 F^{\prime}}$, following the configuration of \cite{Velickovic2017GraphNetworks} with LeakyReLU with a negative slope of $\lambda=0.2$. Pairwise unnormalized attention scores $e_{ij}$ are computed for all nodes $i$ and $j$ that share an edge: 
\begin{equation}
    e_{i j}=\operatorname{LeakyReLU}\left(\mathbf{a}^{\top}\left[ \tilde{\mathbf{h}}_i \| \tilde{\mathbf{h}}_j\right] \right), 
\end{equation}
where $\|$ denotes the concatenation operation of the two linearly transformed node feature vectors $\tilde{\mathbf{h}}_i$ and $\tilde{\mathbf{h}}_j$. The unnormalized attention coefficient $e_{ij}$ indicates the importance of node $j$'s features to node $i$. In order to ensure comparability among neighbors, coefficients are normalized over the neighborhood of node $i$, denoted $\mathcal{N}(i)$:
\begin{equation}
    \alpha_{ij} = \frac{\exp(e_{ij})}{\sum_{k \in \mathcal{N}(i)} \exp(e_{ik})}.
\end{equation}

The resulting normalized attention coefficients $\alpha_{ij}$ are then employed to compute a weighted combination of neighbor features. To stabilize the learning process, we employ multi-head attention with $M=8$ heads, averaging across the per-head outputs before applying the LeakyReLU activation with an identical slope of $\lambda = 0.2$ as before:
\begin{equation}
    \mathbf{h}_i^{\mathrm{GAT}}=\operatorname{LeakyReLU}\left(\frac{1}{M} \sum_{m=1}^M \sum_{j \in \mathcal{N}(i)} \alpha_{i j}^{(m)} \tilde{\mathbf{h}}_j^{(m)}+\mathbf{b}\right).
\end{equation}

Subsequently, neighborhood topologies are refined to account for similarities in the feature space after an initial attention-based message-passing step. This approach aims to enhance the discriminative capabilities along shorelines and over large contiguous surfaces of the same class. This second step is realized through dynamic Edge Convolution (EdgeConv), first introduced by \cite{Wang2019DynamicClouds}. It incorporates the dynamic reassembly of the neighborhood relations based on distance in the feature space, as well as direct modeling of edge relationships. Specifically, following the graph modification, at each EdgeConv layer, for a node $i$, edge features $\mathbf{f}_{ij}$ are constructed for each neighbor $j \in \mathcal{N}^\prime(i)$ in the dynamically redefined neighborhood $\mathcal{N}^\prime(i)$:
\begin{equation}
    \mathbf{f}_{i j}=\mathbf{h}_i^{\mathrm{GAT}} \|\left(\mathbf{h}_j^{\mathrm{GAT}}-\mathbf{h}_i^{\mathrm{GAT}}\right),
\end{equation}
where $\mathbf{h}_i^{\mathrm{GAT}} \in \mathbb{R}^F$ denotes the feature vector of node $i$ after the GATConv step and $\|$ signifies concatenation. Subsequently, a two-layer multilayer perceptron (MLP) with a LeakyReLU activation between its layers is applied to these edge features and aggregated via max-pooling to obtain the updated node representation
\begin{equation}
    \mathbf{h}_i^{\mathrm{Edge}}=\max _{j \in \mathcal{N}^{\prime}(i)} \phi_{\Theta}\left(\mathbf{f}_{i j}\right),
\end{equation}
where $\phi_{\Theta}: \mathbb{R}^{2 F^{\prime}} \rightarrow \mathbb{R}^{F^{\prime}}$ denotes the two-layered MLP parameterized by $\Theta$. 

In our proposed approach, we sequentially combine three Dynamic Attention Graph Convolution (DAGC) blocks, each consisting of a GATConv followed by a dynamic EdgeConv. These blocks are arranged in a progression of increasing layer sizes of 64, 128, and 256. The DAGC blocks are followed by a graph convolutional layer serving as a decoder (GCN Decoder) to obtain node-level prediction labels (see Figure \ref{FIG:model}). The final model architecture consists of approximately 850,000 trainable parameters.

\subsection{Preprocessing and Feature Engineering} \label{sec:feature_engineering}

Each node of a graph is associated with a list of features used as inputs, along with aggregated information from its neighbors. Prior to feature engineering, several preprocessing steps are applied to ensure data integrity and obtain reliable ground truth labels. Subsequently, PIXC pixels are matched spatially to the DSWx-HLS pixels, which corresponds to the assignment of the ground truth label to each SWOT pixel. In this process, PIXCVec coordinates are employed where available to address geolocation errors. PIXC pixels are subsequently matched to DSWx-HLS pixels if the PIXC reference coordinates lie within the DSWx-HLS pixel with a temporal shift of at most 48 hours. DSWx-HLS pixels with moderate or low confidence are excluded to reduce label noise based on the confidence information provided in the \textit{CONF} layer. The resulting matched product takes the shape of the PIXC product, but with an appended ground truth class label from the DSWx-HLS product, and is employed through all stages of the study: training, validation, and testing.

The fundamental principle guiding feature selection is a strategic emphasis on directly measurable PIXC quantities, keeping the feature set closely aligned with the information used by the native PIXC classifier \citep{JPL-D-105504-2023} and establishing a potential baseline for future research. To ensure a consistent level of data integrity, an across-track filter is applied to enforce the intended swath width of 50\,km and nadir gap of 20\,km \citep{Biancamaria2016TheHydrology}, reducing the scene coverage from the nominal 64\,km to the intended 50\,km swath. Additionally, pixels are filtered based on the quality-flag attributes provided in the PIXC product, discarding the two higher levels (bits 16-31) as specified in \citet{swot_swot_2025}. The employed input features are:

\begin{itemize}
    \item Coherent power
    \item Phase noise standard deviation
    \item Height standard deviation within a 50\,m radius
    \item Number of effective medium looks
    \item Prior water occurrence
\end{itemize}

\begin{figure*}
    \centering
	\includegraphics[width=\textwidth]{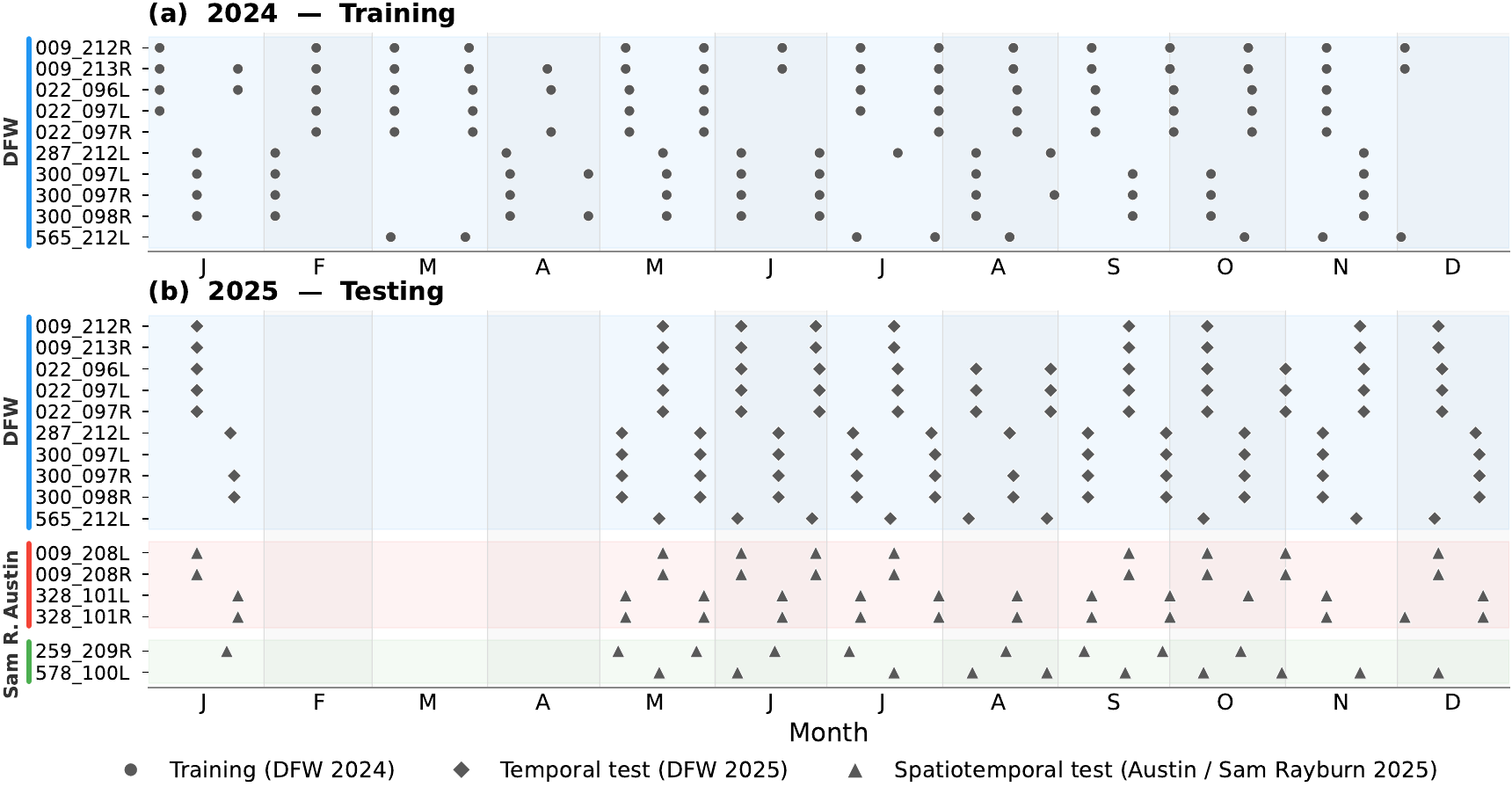}
	\caption{Temporal availability of the SWOT–DSWx-HLS matched dataset. Each marker denotes one matched SWOT–DSWx-HLS acquisition (scene) for a given SWOT tile. Tiles are grouped by region: Dallas–Fort Worth (DFW), Austin, and Sam Rayburn Reservoir (Sam R.). (a) Training data, comprising all 2024 DFW acquisitions (10 tiles, 123 scenes). (b) Independent 2025 test data: DFW tiles to assess temporal generalization (diamonds), while Austin (4 tiles) and Sam Rayburn (2 tiles) assess spatiotemporal generalization (triangles; 109, 42, and 19 scenes, respectively). The model's validation set is a random 20\% patch-level subsample of the 2024 DFW scenes.}
	\label{FIG:temporal_availibility}
\end{figure*}

Coherent power constitutes the primary discriminator, as water surfaces generally return stronger signals due to their specular behavior, whereas land returns are expected to be roughly an order of magnitude lower. Phase noise standard deviation exploits the physical regularity of water surfaces, as water pixels exhibit lower phase noise compared to land (see Section \ref{sec:pixc_data}). The height standard deviation is computed per pixel by considering the per-pixel height attribute provided in the PIXC product within a neighborhood of 50\,m. This is motivated by the notion that water surfaces are typically much flatter than non-water surfaces. The number of effective medium looks provides complementary evidence, as pixels more likely to belong to the water class accumulate more averaged looks during PIXC processing \citep{JPL-D-105504-2023}. Prior water occurrence adds static contextual information at the pixel level. Coherent power, phase noise standard deviation, and height standard deviation all exhibit right-skewed, long-tailed distributions; a log-transform is therefore applied to these features prior to training to enable more efficient learning.

\subsection{Training and Validation}
PIXC tiles considered in this study may contain several million pixels. To reduce computational demands, tiles are split into patches of 2\,km by 2\,km. Each patch is represented as an independent graph, with no edges connecting across patch boundaries. The graph is constructed by linking each node to its ten spatially nearest neighbors, with edges directed from each neighbor toward the node. This is repeated for every node in the patch, such that each node receives exactly ten incoming edges. A graph is considered valid, if it contains at least 100 nodes. As pixel densities vary significantly, the number of nodes per graph may also varies by multiple orders of magnitude. However, as we do not want to exclude any distinct group of ground conditions, we wanted to keep interference in this aspect as minimal as possible. At the same time, we constrain the maximum number of nodes per graph to 10,000 to enable training on a single 24\,GB VRAM GPU.

The dataset is strongly class-imbalanced: water pixels are substantially rarer than land pixels, with an approximate land-to-water ratio of 20:1 across the full dataset. To accommodate this fact, we introduce a series of balancing measures to enable more efficient learning. First, we split the training set into three groups based on the fraction of water nodes in each graph sample. The groups are defined as follows:
\begin{itemize}
    \item Dry: Water fraction < 1\%
    \item Medium: Water fraction 1-10\%
    \item Wet: Water fraction > 10\%
\end{itemize}
Subsequently, we enfore a maximum ratio of three between the most and least represented group. This generally reduces the node-level class imbalance by approximately a factor of three with an acceptable degree of variance across different data splits. Second, we introduce a weight factor that upweighs the minority class in the binary cross-entropy loss. The starting point is the mathematical imbalance factor of the training set. This factor is then downweighted by an empirical factor of 0.8 in favor of slightly higher precision scores (see Equations \ref{eq:precision} and \ref{eq:recall}); this is justified by the fact that, in the face of strong imbalance, very high recall scores can appear frequently without indicating improved model performance. During training, the binary cross-entropy validation loss is minimized. This encourages well-calibrated predictions, as the loss penalizes confident misclassifications more heavily than uncertain ones near the classification threshold. However, this leaves the effective decision boundary produced by the sigmoid function untouched. To achieve better results in the context of targeted applications, such as water level and water surface extent estimation, a balance between precision and recall (see Equation \ref{eq:f1}) is desirable. Consequently, the F1-score-optimal decision boundary is determined on the validation set.

Following this procedure, the model is trained on a total of 10 tiles covering the DFW metropolitan area (see Figure \ref{fig:dallas_tiles}). Training is performed on pixel-wise matched scenes of PIXC and DSWx-HLS data (see Section \ref{sec:opera}). The total stratified training dataset comprises more than 50,000 disconnected graphs or patches. An 80-20 data split is applied to construct training and validation sets.

The selected model's performance is validated on temporally independent data: while training is performed exclusively on scenes acquired in 2024, all assessments are performed on scenes from 2025. We perform temporal generalization tests on the same list of tiles used during training, focused on the DFW metropolitan area (blue in Figure \ref{fig:dallas_tiles}), to assess temporal generalizability. Moreover, we perform spatial generalization tests on unseen tiles over the area of the Sam Rayburn Reservoir in eastern Texas, as well as tiles focused on the city of Austin. The spatial generalization test tiles are visualized in red and green, respectively, in Figure \ref{fig:dallas_tiles}. All tests are performed on labeled data, i.e., they are assessed based on performance metrics with respect to ground-truth labels based on the matched data. This further implies that only scenes and pixels are considered for which matches between SWOT acquisitions and DSWx-HLS were possible based on the criteria detailed in Section \ref{sec:feature_engineering}.

The classification performance is assessed based on the following three metrics:

\begin{align}
    \text{Precision} &= \frac{TP}{TP + FP} , \label{eq:precision} \\
    \text{Recall} &= \frac{TP}{TP + FN}, \label{eq:recall} \\
    F_1 &= 2 \cdot \frac{\text { Precision } \cdot \text { Recall }}{\text { Precision }+ \text { Recall }}, \label{eq:f1}
\end{align}
where TP and FP refer to the total number of true positives and false positives, respectively. Analogously, TN commonly refers to the number of true negatives and FN to the number of false negatives. In this context, a \textit{negative} corresponds to a non-water (or land) prediction, while a \textit{positive} is a water pixel. Each metric qualifies for inclusion due to its specific characteristics. Precision measures the proportion of correctly predicted positive instances (water pixels), while recall measures the proportion of true water pixels successfully identified. In applications for which classified SWOT data may be used, precision is relevant to both surface-area and water-level estimation, while the recall score is primarily relevant to surface-area estimation. Ultimately, in surface area estimations, both metrics are critical since only true water instances shall be classified as such (perfect precision), and simultaneously, all of them shall be captured (perfect recall). Consequently, we employ the F1 score as the primary performance metric in this study, as it represents the harmonic mean of precision and recall and therefore provides a balance between the implications of the respective metrics for downstream applications.

In order to compute the above-listed scores and facilitate comparison with the proposed model, the seven-class classification provided in the PIXC product is mapped to a binary water/non-water classification. In this, all water-associated classes, namely ``water near land'', ``open water'', ``dark water'', ``low-coherence water near land'', and ``open low-coherence water'' (indices 3-7) are mapped to the \textit{water} class, and ``land'' and ``land near wate'' (indices 1 and 2) are mapped to the \textit{non-water} or \textit{land} class.

\section{Results and Analysis} \label{sec:results}

\subsection{Temporal Generalization Performance}

This subsection reports classification performance on the temporal test set, beginning with overall metrics before turning to class-specific and spatial patterns.

\subsubsection{Overall Performance}

The temporal test set encompasses 109 scenes acquired in the year 2025 (see Figure \ref{FIG:temporal_availibility}). The test data were retrieved in a manner consistent with the training set, covering the same set of 10 tiles across the DFW metropolitan area, depicted in Figure \ref{fig:dallas_tiles}a. In the following, we refer to the DAGC-based modeling approach developed in this study as the GNN classifier, and designate the binary class mapping of the original seven SWOT class labels, which are already contained in the PIXC product, as the SWOT baseline classifier, in figures and tables abbreviated as SWOT. This nomenclature will be used throughout the remainder of this paper.

As demonstrated in Table \ref{tbl1}, the GNN classifier predictions exhibit a mean F1 score of 0.86 across all scenes, relative to the binary class labels from the DSWx-HLS-based ground truth. With a standard deviation of 0.20, it also denotes a smaller degree of variation, in addition to the distinctly higher level of agreement compared to the binary SWOT classifier's F1 score of 0.52$\pm$0.24. The GNN classifier's median F1 score of 0.93 is notably higher than its mean, indicating a left-skewed distribution across scenes, a phenomenon that is not observed for the SWOT classifier.

\begin{table}
\caption{Classification performance of the developed GNN model and the binary SWOT classifier on the temporal test set}\label{tbl1}
\begin{tabular*}{\tblwidth}{@{}LLL@{}}
\toprule
Metric & GNN classifier & SWOT binary classifier\\ 
\midrule
Median F1      & $\mathbf{0.93}$        & $0.52$ \\
Mean F1 (std)        & $\mathbf{0.86} \,\,(0.20)$ & $0.52 \,\,(0.24)$ \\
Mean precision (std) & $\mathbf{0.88} \,\,(0.19)$ & $0.42 \,\,(0.24)$ \\
Mean recall (std)    & $\mathbf{0.86} \,\,(0.20)$ & $0.83 \,\,(0.18)$ \\
\bottomrule
\end{tabular*}
\end{table}

\begin{figure}
	\centering
	\includegraphics[width=\linewidth]{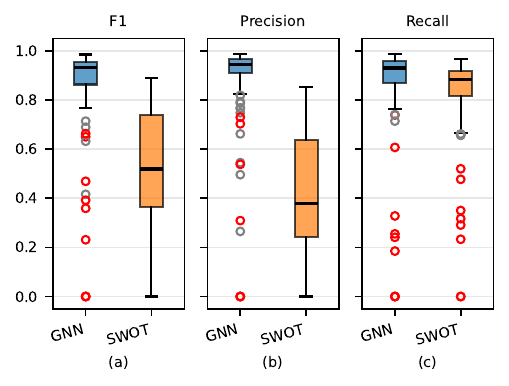}
	\caption{Temporal generalization performance of the GNN model vs. the SWOT baseline classifier's class labels (PIXC contained class labels). Outliers in red: scenes that contain fewer than 250 patches.}
	\label{FIG:temporal_model_comparison_boxplots}
\end{figure}

As illustrated in Figure \ref{FIG:temporal_model_comparison_boxplots}, the temporal model comparison boxplots demonstrate that the improvements persist across all three metrics. However, the nature of the improvement differs by metric. The GNN classifier's interquartile range (IQR) for the F1 score is notably more confined and elevated (0.83 to 0.95) compared to the SWOT classifier's range (0.37 to 0.74). The precision score indicates a more pronounced contrast between the models, with the GNN classifier's IQR ranging from 0.91 to 0.97, in contrast to the SWOT classifier's considerably broader and lower IQR of 0.24 to 0.64. Finally, the recall score,  illustrated in Figure \ref{FIG:temporal_model_comparison_boxplots}c, denotes more similar IQRs, ranging from 0.87 to 0.96 and from 0.82 to 0.92 for the GNN- and SWOT classifiers, respectively.  Numerically, the recall improvement is comparatively subtle compared to the substantial gains in precision. As demonstrated in Table \ref{tbl1}, both patterns are also reflected in the mean values. The found asymmetry in improvements across precision and recall scores can be attributed to the false-positive-dominated error pattern in the SWOT classifier, consistent with the bright land misclassification mechanism. As a result of the bright land error pattern, the SWOT classifier denotes increased numbers of false positives, which, when improved, directly impact the precision score through the decreased contribution to the denominator (see Equation \ref{eq:precision}). The recall score naturally benefits less from this same mechanism, as it is sensitive to a change in false negatives.

There are a number of outlier scenes, as defined by the 1.5$\cdot$IQR rule, found in Figure \ref{FIG:temporal_model_comparison_boxplots}. We find 69\% of F1 score outliers and 82\% of recall outliers for the GNN classifier, coinciding with low patch counts of less than 250 patches per scene. Precision score outlier scenes denote a lower share of 41\% of precision, potentially indicating an additional source of error for precision that is not explained by patch count alone. The SWOT classifier exhibits no outliers for the F1 score and precision, which is a consequence of its wider interquartile range rather than higher consistency.

\subsubsection{Class-specific Error Patterns}

\begin{figure}
	\centering
	\includegraphics[width=\linewidth]{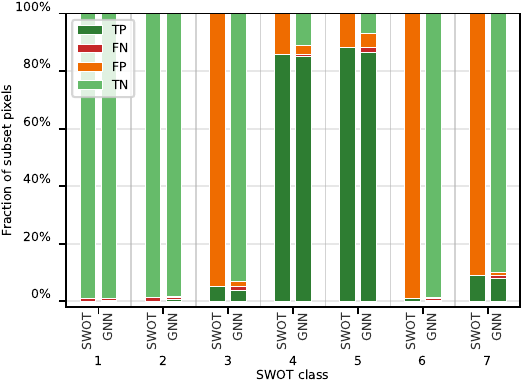}
	\caption{Confusion decomposition of the binary classification results per 7-class SWOT classification on the temporally disconnected test set. The seven classes are 1: land, 2: land near water, 3: water near land, 4: open water, 5: dark water, 6: low-coherence water near land, 7: open low-coherence water.}
	\label{FIG:temporal_model_comparison_confusion}
\end{figure}

Having established the overall gains in binary water-land classification, the subsequent focus is on the origin of these gains. The confusion analysis per original SWOT class (seven-class classification), as illustrated in Figure \ref{FIG:temporal_model_comparison_confusion}, offers a comprehensive view. As the SWOT binary classifier assigns a single predicted label to every pixel within a given original SWOT class, it cannot produce false negatives or true negatives for the water-related classes (indices 3-7), nor true positives or false positives for the land-related classes (indices 1-2). The GNN classifier, on the other hand, directly predicting a binary class label per pixel, is not subject to this constraint.

\begin{figure}
	\centering
	\includegraphics[width=\linewidth]{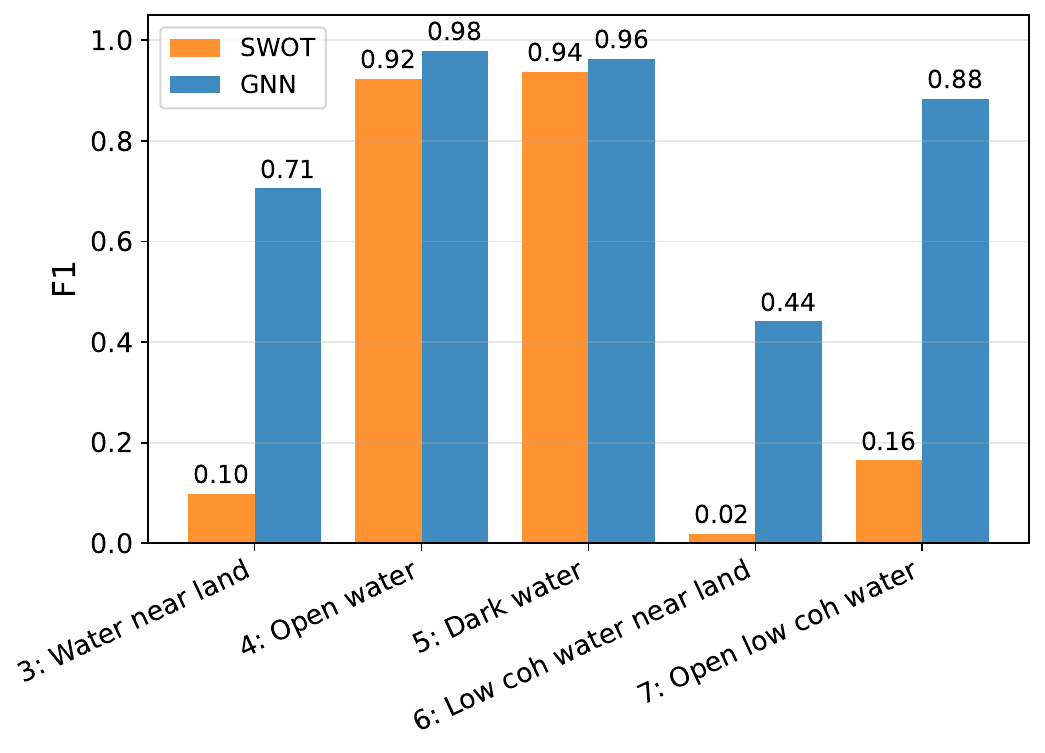}
	\caption{F1 scores of binary class predictions per selected SWOT class labels on the test set for the original SWOT classifier on the proposed model (GNN)}
	\label{FIG:temporal_model_comparison_f1}
\end{figure}

For the \textit{land} and \textit{land near water} classes, both classifiers achieve a large share of over 98\% correctly predicted class labels, with a slight advantage to the GNN classifier. This advantage is primarily driven by a reduced number of false negatives compared to the SWOT baseline. Moreover, for the two primary classes not adjacent to shorelines or labeled as being subject to low coherence, \textit{open water} and \textit{dark water}, the SWOT classifier already performs distinctly better than for the two near-shore classes, \textit{water near land} and \textit{low-coherence water near land}, and the \textit{open low-coherence water} class. Comparing the SWOT to the GNN classifier for the \textit{open water} and \textit{dark water} classes, the overall dominant pattern is a reduction in the share of false positives. The fraction of true positives remains essentially unchanged for \textit{open water} and decreases modestly for \textit{dark water} under the GNN classifier. This denotes a minor exception and remains without an impact on the F1 scores, as enhancements from 0.92 to 0.98 and from 0.94 to 0.96 are denoted for \textit{open water} and \textit{dark water}, respectively (see Figure \ref{FIG:temporal_model_comparison_f1}). We find this modest decrease in true positive detections, w.r.t. to the SWOT classifier, likely to be attributable to the feature similarity of dark water pixels to land. In instances where larger patches of dark water are present, the neighborhood-based aggregation mechanism of the DAGC-based GNN model has limited corrective capabilities, as the surrounding pixels are similarly dark rather than providing a contrasting signal. A larger neighborhood definition could prove to be beneficial under these circumstances.

Analyses of the near-shore and low-coherence classes reveal a consistent, more pronounced version of the dominant pattern found for the \textit{open water} and \textit{dark water} classes above. According to the ground truth, 95\% of pixels labeled \textit{water near land}, 91\% of pixels labeled \textit{open low-coherence water}, and nearly 99\% of pixels labeled \textit{low-coherence water near land} belong to the \textit{land} class, a systematic disagreement between the SWOT-native classification and the DSWx-HLS product. These pixels represent false positives, which we find to be predominantly caused by bright land, consistent with the phenomenological similarity between bright-land occurrences and water. The three labels are frequently assigned when the bright-land ambiguity arises, rather than reflecting a true extension of water extent along actual shorelines. This is reflected in the poor SWOT F1 scores of 0.10, 0.16, and 0.02 for \textit{water near land}, \textit{open low-coherence water}, and \textit{low-coherence water near land}, respectively (see Figure \ref{FIG:temporal_model_comparison_f1}).

\begin{figure*}
	\centering
	\includegraphics[width=\textwidth]{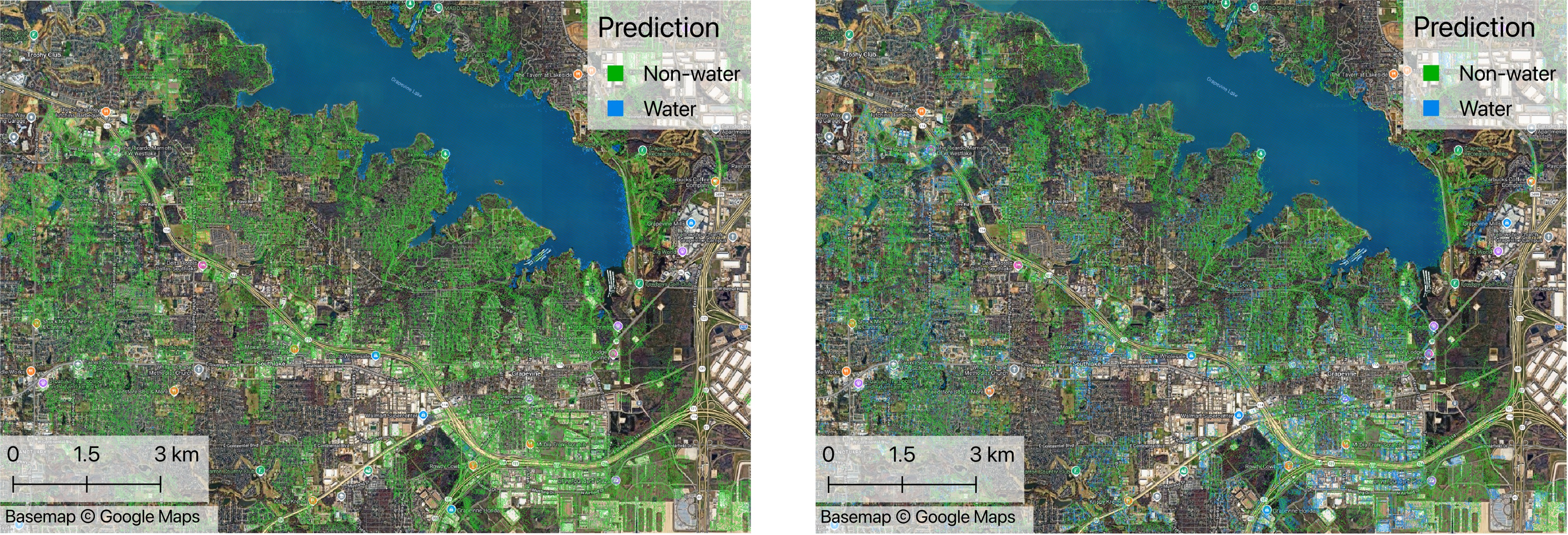}
	\caption{Comparison of the GNN-based binary classification result (left) and the native SWOT classifier (right) over Grapevine Lake with the northern end of DFW International Airport to the southeast (022\_097L on 2025-08-31)}
	\label{FIG:grapevine_lake}
\end{figure*}

The GNN classifier demonstrates the capacity to learn and correct this pattern. Within each of the three classes, it identifies approximately 98\% or more of the ground-truth-land pixels as true negatives, and the effect is clearly visible in Figure \ref{FIG:grapevine_lake}, where the systematic false-positive pattern of the SWOT classifier is substantially reduced. This transition from false positives to true negatives is the primary factor behind the improvements in these classes, mirroring at a larger scale the shift already noted for \textit{open water} and \textit{dark water}. The degree of correction nonetheless varies with the original SWOT class. The \textit{low-coherence water near land} class remains the most challenging (F1: 0.44, 
$\Delta$F1: 0.42), followed by \textit{water near land} (F1: 0.71, $\Delta$F1: 0.61). \textit{Open low-coherence water} denotes the largest, absolute F1 score improvement with a $\Delta$F1 of 0.72 as the GNN achieves an F1 of 0.88.

\subsubsection{Spatial and Water-fraction Dependence}

\begin{figure*}
	\centering
	\includegraphics[width=0.8\textwidth]{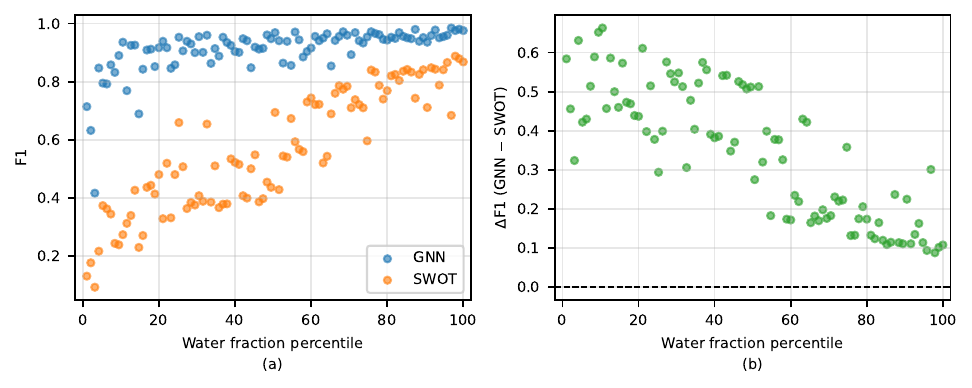}
	\caption{Temporal generalization performance of the GNN model and the SWOT baseline class labels in terms of F1 scores per water fraction percentile (a) and the relative F1 score improvements of the GNN compared to the binary SWOT class labels (b). The water fraction quantiles are defined as follows and were derived based on the binary ground truth class labels: p25:0.0514; p50: 0.0695; p75: 0.1065. Only scenes containing a minimum number of 250 patches are visualized.}
	\label{FIG:temporal_f1_vs_water_fraction_percentile}
\end{figure*}

The temporal generalization performance of the GNN model and the SWOT baseline classifier is presented in terms of F1 scores per water fraction percentile in Figure \ref{FIG:temporal_f1_vs_water_fraction_percentile}a, with the corresponding relative F1 improvements of the GNN over the binary SWOT class labels shown in Figure \ref{FIG:temporal_f1_vs_water_fraction_percentile}b. The water fraction percentiles are derived from the binary ground truth labels, with the 25th percentile (p25) at 0.0514, the 50th percentile (p50) at 0.0695, and the 75th percentile (p75) at 0.1065. The improvement is largest under dry, land-dominated conditions and diminishes as scenes grow wetter. Near the 20th percentile, $\Delta$F1 reaches approximately 0.5, then declines in a continuous though noisy manner to roughly 0.10 to 0.15 by the 80th percentile, before plateauing.

\begin{table}
\caption{List of the temporal test tiles with the mean F1 score achieved by the GNN classifier for each tile for the test period of 2025. The \textit{Scenes} column quantifies the number of labeled scenes that are contained in the dataset, and the column \textit{< 250} indicates how many of those labeled scenes contain fewer than 250 patches.}\label{tbl2}
\begin{tabular*}{\tblwidth}{@{}LLLLL@{}}
\toprule
Tile  & Scenes & < 250 & Mean F1  & Mean F1 (filtered)\\ 
\midrule
022\_096L & 12 & 0 & \textbf{0.95} & 0.95\\
009\_213R & 9 & 1 & 0.94 & 0.94\\
300\_097L & 10 & 1 & 0.90 & \textbf{0.96}\\
300\_097R & 12 & 0 & 0.90 & 0.90 \\
565\_212L & 9 & 2 & 0.88 & 0.90 \\
009\_212R & 9 & 2 & 0.87 & 0.88 \\
022\_097L & 12 & 1 & 0.85 & 0.90 \\
287\_212L & 12 & 1 & 0.84 & 0.90\\
022\_097R & 12 & 2 & 0.76 & 0.84 \\
300\_098R & 12 & 4 & 0.75 & 0.94\\
\bottomrule
\end{tabular*}
\end{table}

\begin{figure*}
	\centering
	\includegraphics[width=.8\textwidth]{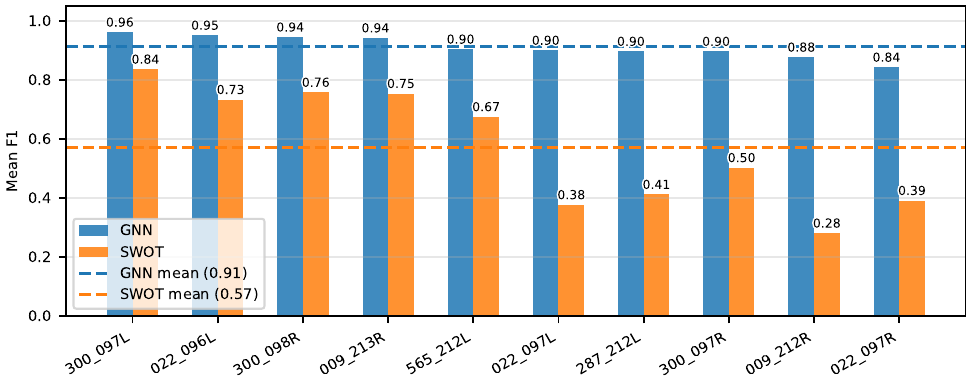}
	\caption{Full-scene mean F1 scores per tile and model on the temporal generalization test set (fewer than 250 patches/scene were filtered out).}
	\label{FIG:temporal_f1_per_location}
\end{figure*}

The per-location mean F1 scores are summarized in Table \ref{tbl2}, both including and excluding scenes with fewer than 250 patches. We find that the relative ordering of tile-level performance is sensitive to the inclusion of low-patch-count scenes, with the unfiltered mean F1 score ranging from 0.75 to 0.95 across tiles, compared to 0.84 to 0.96 after filtering scenes with fewer than 250 patches. The filtering also has effects on the overall means, resulting in an increase from 0.86 to 0.91. The full-scene mean F1 scores per tile and model on the temporal generalization test set are presented in Figure \ref{FIG:temporal_f1_per_location}. The GNN classifier achieves a mean F1 score of 0.91 across all tiles, compared to 0.57 for the SWOT baseline. The largest enhancements are noted for tiles 022\_097L, 022\_097R, and 009\_212R, which overlap significantly in an area with a considerable presence of bright land surrounding the core of the DFW metropolitan area. A visual representation of the GNN-based and SWOT-based binary classification results for this area is provided in Figure \ref{FIG:grapevine_lake}, where the SWOT-based classification displays a systematic pattern of erroneous land classification in the highly urbanized area extending from the southern end of Grapevine Lake to the northern end of the DFW International Airport in the southeast.

\subsection{Spatiotemporal Generalization Performance}

Beyond temporal generalization, the classifier's transfer capabilities across geographical locations are assessed. This spatiotemporal testing procedure was conducted on 61 labeled scenes in the Austin area (tiles 009\_208L, 009\_208R, 328\_101R, and 328\_101L) and the Sam Rayburn Reservoir (tiles: 578\_100L, 259\_209R), see Figures \ref{fig:dallas_tiles} and \ref{FIG:temporal_availibility}. The findings indicate a moderate decline in the GNN classifier's performance on this spatiotemporal test set relative to its performance on the temporally-disconnected test set (Table \ref{tbl1}), with the median F1 score decreasing by 8.6\%. Analogously, we also note a decrease in the mean F1 score of 14.0\%. The detailed results are outlined in Table \ref{tbl3}. A decline is expected when transitioning to geographically different locations, as the learned discrimination capabilities are linked to the surface conditions encountered during training. Consequently, direct comparisons across test sets may not be significant. To contextualize the magnitude of these declines, it is beneficial to consider them relative to the underlying advantage of the GNN classifier over the SWOT baseline within each test set. On the temporal test set, the GNN classifier's median F1 score exceeds that of the SWOT classifier by 0.41 (a relative improvement of 80.0\%), while on the spatiotemporal test set this absolute margin remains comparable at 0.40, corresponding to an even larger relative improvement of 89.8\%. A similar pattern holds for the mean F1 score, where the GNN's absolute gain over SWOT is 0.34 on the temporal test set and 0.31 on the spatiotemporal test set, with the relative improvement increasing from 64.3\% to 72.1\%. The observed declines in GNN performance when moving to the spatiotemporal test set are therefore modest in comparison to the sustained, and in relative terms even strengthened, advantage of the GNN classifier over the SWOT baseline. 

\begin{table}
\caption{Classification performance of the developed GNN model and the binary SWOT classifier on the spatiotemporal test set}\label{tbl3}
\begin{tabular*}{\tblwidth}{@{}LLL@{}}
\toprule
Metric & GNN classifier & SWOT binary classifier\\ 
\midrule
Median F1      & $\mathbf{0.85}$        & $0.45$ \\
Mean F1 (std)        & $\mathbf{0.74}\pm0.29$ & $0.43\pm0.33$ \\
Mean precision (std) & $\mathbf{0.72}\pm 0.31$ & $0.35\pm 0.32$ \\
Mean recall (std    & $\mathbf{0.80}\pm 0.24$ & $0.76\pm 0.22$ \\
\bottomrule
\end{tabular*}
\end{table}

\begin{figure}
	\centering
	\includegraphics[width=\linewidth]{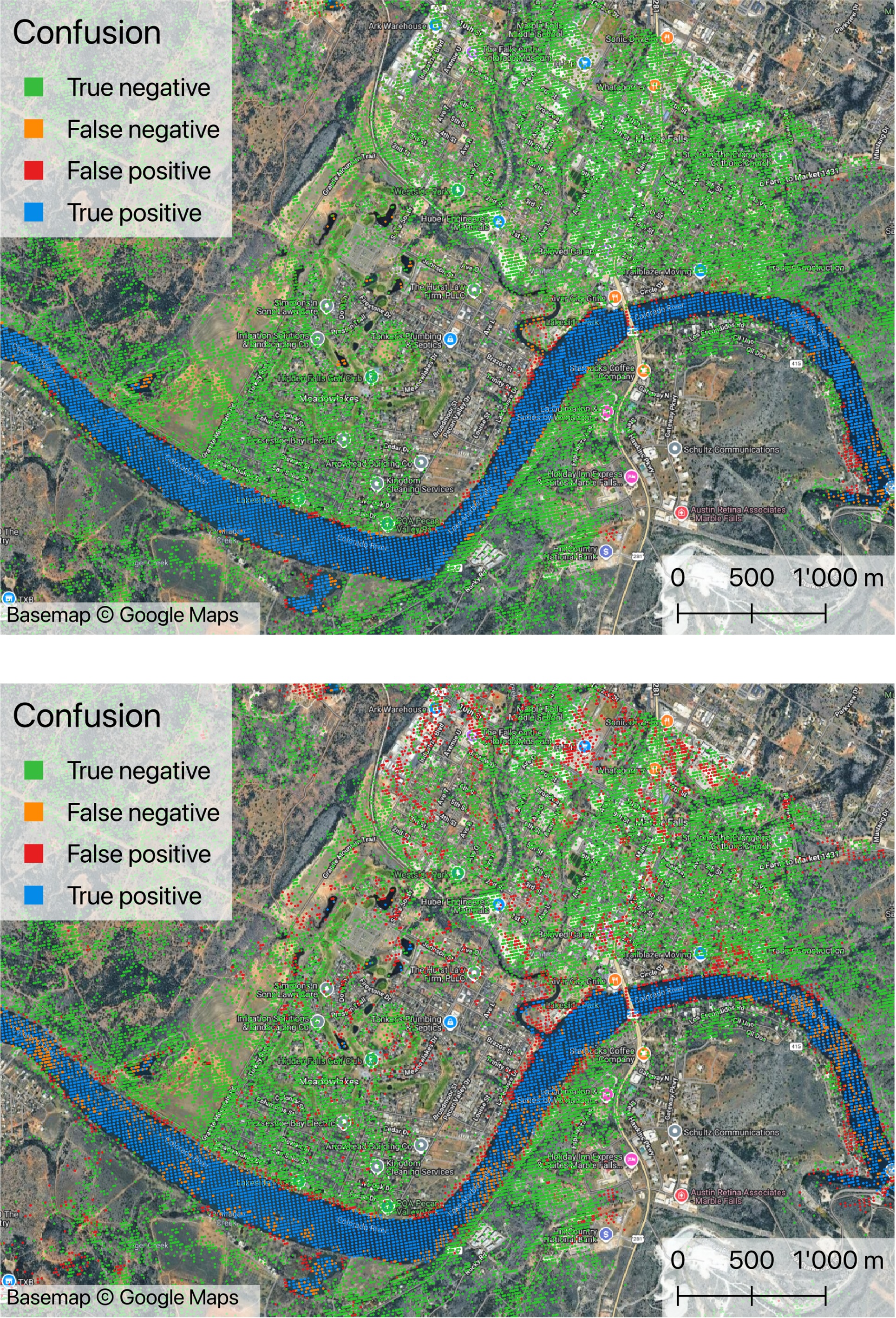}
	\caption{Comparison of the binary water-land classification of the GNN-based classifier (top) and native SWOT PIXC classifier (bottom) along the Colorado River near Marble Falls, TX (328\_101R on 2025-09-11).}
	\label{FIG:colorado_river}
\end{figure}

Riverine structures that meet the SWOT science requirement for a minimum width of 100\,m were lacking in the training set, which was predominantly composed of lakes and reservoirs. The results presented in Figure \ref{FIG:colorado_river} therefore serve as a qualitative evaluation of the classifier's capacity to generalize to a range of distinct water body types. A primary enhancement is the effective reduction of bright land effects at the scene level. Additionally, it exhibits a decrease in class-label noise along shorelines and across the entire river structure, particularly in the narrower sections of the investigated segment. The reduced noise combined with the model's apparent ability to capture the sharp boundary of the shoreline indicates that the chosen design can plausibly resolve sharp class boundaries. For very small water bodies situated centrally within the depicted scene as part of a golf course, measuring approximately 50 to 60\,m or smaller, the model's predictions are predominantly \textit{land}, leading to false negatives. This observation is consistent with the over-smoothing commonly observed in deeper GNNs \citep{Li2018DeeperInsights}. In such instances, the SWOT classifier can detect water bodies, but it also generates a substantial number of false positives around these water features.

It is worth noting that this shoreline analysis is subject to some degree of uncertainty, as the DSWx-HLS-based ground truth is itself limited to its 30\,m spatial resolution in fine-scale boundary delineation; some of the observed improvement may reflect alignment with ground-truth boundary artifacts rather than the true shoreline, a deviation that cannot be further quantified with the present data. Accurate classification at the finest PIXC spatial resolution therefore remains a methodological challenge rather than a solved problem.

\subsection{Sensitivity to Geolocation} \label{sec:geolocation_sensitivity}

\begin{figure*}
    \centering
    \includegraphics[width=0.6\textwidth]{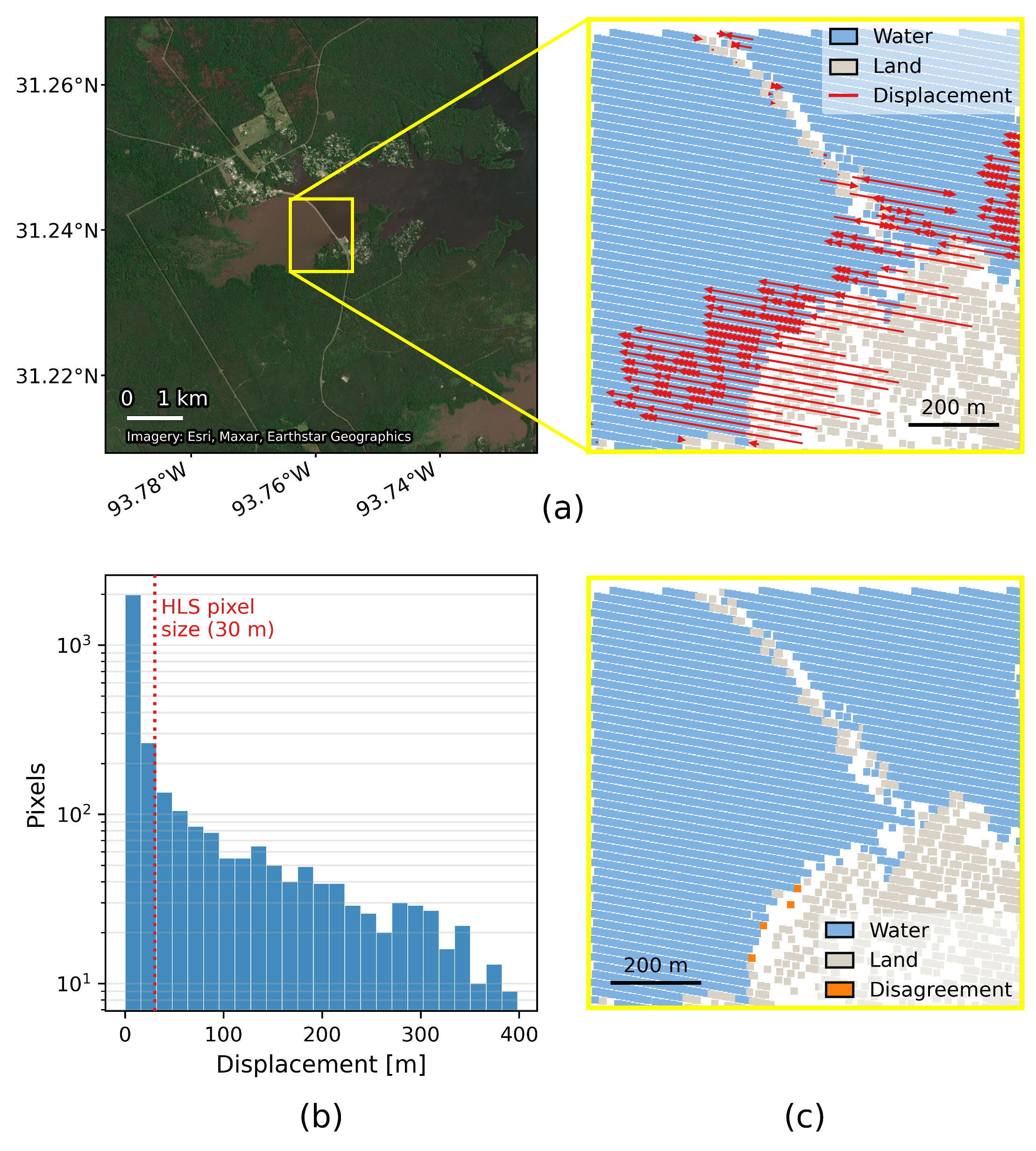}
	\caption{Example from the Sam Rayburn Reservoir test scene (259\_209R; acquired on 2025-10-21). Panel (a) shows geolocation errors in the form of displacement vectors from native PIXC to PIXCVec-corrected pixel positions, overlaid on the model's predicted classification. Panel (b) shows the distribution of geolocation error magnitudes for the pixels shown in (a), with the spatial resolution of the DSWx-HLS reference product marked in red. Panel (c) shows pixels where predictions differ between the models trained on data with and without the PIXC + PIXCVec geolocation correction.}
	\label{FIG:geolocation_issue}
\end{figure*}

We compare model performance with and without the PIXCVec-based geolocation correction described in Section \ref{sec:feature_engineering}. In all cases, models are evaluated against the same geolocation-corrected test set. Differences in performance therefore only reflect the differences in the training data, caused by the differing handling of the geolocation attributes. Additional details on the model trained without geolocation correction are provided in Appendix \ref{sec:pixc_model}.

At small spatial scales, the effect of the geolocation correction can be substantial. Figure \ref{FIG:geolocation_issue}a illustrates that the correction frequently changes a pixel’s ground truth label. These label changes reflect spatial displacements that can reach several hundred meters between native PIXC and PIXCVec-corrected positions (Figure \ref{FIG:geolocation_issue}b). At the scene level, however, performance metrics remain unaffected, as shown by model performances across the test set employed in this study (Tables \ref{tbl1}, \ref{tbl3}, and \ref{tbl4}). Only pixels near water bodies are present in the prior databases and are therefore eligible for a PIXCVec correction. Within this subset, only a fraction shift far enough to change labels. In the close-up shown in Figure \ref{FIG:geolocation_issue}c, the two models disagree on only four pixels, with metrics unchanged at F1 = 0.96, precision = 0.95, and recall = 0.97.

\section{Discussion} \label{sec:disc}

The results presented in Section \ref{sec:results} demonstrate that the primary source of the GNN classifier's advantage over the native SWOT binary classifier is the suppression of bright land false positives, with the largest gains observed in scenes with low water fractions. Low-water-fraction scenes are naturally dominated by land surfaces and can therefore represent the conditions that give rise to the bright land ambiguity underlying the dominant error sources in SWOT class labels. As established in the previous section, near-shore and low-coherence class labels are in large numbers assigned to bright land pixels. This has a direct consequence for how the reported gains should be interpreted. The aggregate F1 scores in Table \ref{tbl1} average over both low- and high-water-fraction scenes, and therefore blend the largest gains together with smaller ones. Applications concerned specifically with water detection in urban areas, including shoreline mapping, flood-extent delineation, and reservoir-boundary tracking, predominantly operate under low-water-fraction conditions. Such applications would therefore be expected to benefit from gains closer to the upper end of this range.

While the improvements across these classes are driven largely by the GNN's ability to separate bright land from water, the performance differences that remain are themselves informative about the underlying pixel characteristics. The two near-shore classes perform worst, likely because some of their pixels are located along shorelines rather than being tied solely to bright land misclassification, which makes them intrinsically harder to classify. Based on the reasonably strong F1 score of 0.71 for the \textit{water near land} class, the architectural mechanism of balancing feature similarity and spatial proximity appears to be a suitable mitigation measure. Performance declines again under low coherence, as seen in the \textit{low-coherence water near land} class. We attribute the superior performance of \textit{open low-coherence water} to its position away from shorelines, where it frequently occurs as isolated pixels, allowing the GNN to draw on nearby open-water pixels to offset low-coherence noise.

The question of whether this advantage persists beyond the geographic target area included during training is directly addressed by the spatiotemporal test results. They show that the GNN classifier retains a pronounced advantage under geographic transfer. Although performance declines relative to the temporal test set, this decline understates how well the model generalizes. The absolute gap in F1 scores between the two classifiers remains stable between the temporal and spatiotemporal test sets, and the relative improvement over SWOT increases in both cases. The advantage of the GNN classifier over the bright land errors that dominate SWOT's near-shore and low-coherence labels is therefore not confined to the tiles and conditions encountered during training. This robustness has value beyond the present study, since performance that survives geographic extrapolation without retraining is a prerequisite for extending the approach toward broader-scale models, where exhaustive region-specific training data comes with significant effort. The evaluated test sets, however, do not cross major climatic boundaries, and the findings cannot be assumed to extend to fundamentally different regimes such as tropical floodplains or high-latitude environments, where a global model would eventually need to perform.

The strengths and limitations of the proposed model follow directly from its mechanism. The model is most effective where bright land misclassification dominates the native SWOT labels, which is why the gains are largest for the near-shore and low-coherence classes and more modest for open water and dark water, where SWOT is already reliable. This targeted effectiveness comes with well-characterized bounds. The clearest is the smoothing trade-off that suppresses small, isolated water bodies of approximately 50 to 60\,m, as seen in the Colorado River evaluation, where the same neighborhood-based aggregation that removes bright-land false positives can also attenuate a weak but genuine water signal. A second limitation follows from the binary formulation itself, which marks detected pixels as water without further differentiation. This is suitable for extent estimation but does not by itself yield reliable height information required for water-level or discharge retrieval. A further consideration concerns the geolocation noise found in PIXC data. Using PIXCVec-corrected geolocation for the label assignment of the training does not measurably change aggregate classification performance relative to native PIXC geolocation, despite substantial pixel displacements. This indicates that the reported gains are independent of the geolocation noise. Applications relying on precise pixel geolocation near small or narrow water bodies, such as shoreline mapping and boundary-focused tasks, however, remain sensitive to this geolocation uncertainty, motivating the combination of PIXC and PIXCVec. 

The computational demands differ substantially between training and inference. The training process of the proposed model required approximately four days on a single NVIDIA RTX 4090 GPU with 24\,GB of memory. Inference on an average-sized scene from the test set (009\_208L\_2025-11-02; approximately 2.62 million pixels after preprocessing) completed with a compute time of slightly over two minutes, and a peak GPU memory usage of 2,771\,MB. The largest scene in the test set (022\_097L\_2025-06-30; approximately 6.4 million pixels), for comparison, required slightly over five minutes and 3,808 MB of peak GPU memory on the same GPU. Therefore, despite the training process requiring several days, the application of the model to individual scenes is accomplished within a few minutes, exhibiting significantly reduced GPU memory requirements, making it suitable to be employed on a wide range of consumer-grade GPUs.

The findings, when considered collectively, establish pixel cloud-level classification with the proposed approach as an effective step toward enhanced water detection directly in SWOT pixel clouds, most immediately for boundary-focused extent mapping, with small-water-body recovery mechanisms and height retrieval marking the principal directions for further development.

\section{Conclusion and Outlook}

In this study, we have developed a graph neural network approach for water detection that operates directly at the PIXC level. It employs an innovative combination of attention-based feature aggregation and dynamic graph adaptation based on feature similarity in a unified block structure. The model was trained on a full year of PIXC data over the DFW metropolitan area with DSWx-HLS labels spatiotemporally matched to the combined PIXC and PIXCVec pixels to serve as ground truth. In our evaluations, we find that the model substantially outperforms the native PIXC classifier in terms of binary classification on both the temporal and spatiotemporal test set. In our evaluations, we find that the model significantly outperforms the native PIXC classifier in terms of binary classification, with an achieved mean scene-level F1 score of 0.86 compared to 0.52 by the native classifier on the temporally disconnected test set. Furthermore, we also identify promising spatial generalizability when transferring to different ground conditions achieving an average F1 score improvements of 0.31 to achieve a mean F1 score of 0.74; however, without crossing any major climatic boundaries.

From the findings presented in this article, we conclude that the proposed modeling approach enables effective water detection in SWOT pixel clouds, reducing misclassifications caused by bright land from built surfaces. With this approach, we are able to achieve high-quality water detection natively on PIXC scenes, thanks to the graph architecture, allowing for the maximal spatial resolution offered by Level-2 SWOT products, with PIXC geolocation noise having virtually no effect on overall model performance. The proposed model achieves sharp land-water delineation that is in high agreement with the DSWx-HLS product used as reference, given that the water bodies are of a sufficient size. For very small water bodies that also do not lie within the initial SWOT science requirements, the model approach tends to smooth them out and miss them. This limitation, found in the current implementation, presents a highly relevant avenue for future research.

Overall, we consider the presented approach a promising step toward more reliable water detection from SWOT PIXC. The model already enables reliable surface extent quantification, and with adequate denoising strategies or stricter quality control, it could enable broader water monitoring applications involving water height measurements in the study region, including urban flood monitoring. Investigating the latter offers a natural avenue for future research, with the objective of a detailed assessment of model behaviour under substantial deviations from prior water occurrence information, which could not be evaluated using the available data in the framework of this study. Furthermore, extending the approach to additional climatic regions would help assess its potential as a resource-efficient scaling strategy for future global models, as suggested by the generalization capabilities demonstrated in this study.

\section*{Data Availability}
The SWOT Level 2 Water Mask Pixel Cloud data (SWOT\_L2\_HR\_PIXC, Version D) used in this study are openly available from the NASA Physical Oceanography Distributed Active Archive Center (PO.DAAC) at \url{https://doi.org/10.5067/SWOT-PIXC-D}, alongside its auxiliary data product, SWOT\_L2\_HR\_PIXCVec\_D, at \url{https://doi.org/10.5067/SWOT-PIXCVEC-D}. The OPERA Dynamic Surface Water Extent from Harmonized Landsat Sentinel-2 data (OPERA DSWx-HLS, Version 1.0) used in this study are openly available at \url{https://doi.org/10.5067/OPDSW-PL3V1}.

\section*{Code Availability}
The primary code used for data processing, model training and inference, as well as the trained model weights are available from the corresponding authors upon reasonable request.

\printcredits

\section*{Declaration of Competing Interests}
The authors declare no competing interests.

\bibliographystyle{cas-model2-names}

\bibliography{cas-refs}


\pagebreak

\appendix

\setcounter{table}{0}
\renewcommand{\thetable}{A.\arabic{table}}

\section{PIXC Model} \label{sec:pixc_model}
As referenced in Section \ref{sec:geolocation_sensitivity}, we provide an additional model, which was trained purely on PIXC data without any additional prior information from PRD/SWORD or PLD. The model architecture and training strategy are identical to that of the main model described in Section \ref{sec:themodel}. The only difference lies in the geolocation information, which, for this model, is solely based on PIXC, rather than a fusion of PIXC and PIXCVec. In Table \ref{tbl4}, we provide a comparison on the PIXCVec-corrected test set between the proposed GNN model (referred to as the main model) and the PIXC model. 

\begin{table*}
\caption{Performance of the PIXC model and proposed main model (trained on PIXC + PIXCVec) on the PIXCVec-corrected 2025 test set, for scenes with at least 250 patches}\label{tbl4}
\begin{tabular*}{\textwidth}{@{}LLLLLLL@{}}
\toprule
Test Set & Model & Scenes & Precision & Recall & F1 (mean) & F1 (median)\\ 
\midrule
Temporal       & PIXC & $95$  & $0.91\pm0.11$ & $0.92\pm0.06$ & $0.91\pm0.08$ & $0.94$ \\
               & main & $95$  & $0.92\pm0.11$ & $0.91\pm0.06$ & $0.91\pm0.08$ & $0.94$ \\
\midrule
Spatiotemporal & PIXC & $63$  & $0.74\pm0.23$ & $0.78\pm0.21$ & $0.74\pm0.22$ & $0.81$ \\
               & main & $63$  & $0.75\pm0.23$ & $0.77\pm0.22$ & $0.74\pm0.22$ & $0.82$ \\
\midrule
All            & PIXC & $158$ & $0.84\pm0.19$ & $0.86\pm0.16$ & $0.84\pm0.17$ & $0.91$ \\
               & main & $158$ & $0.85\pm0.19$ & $0.85\pm0.16$ & $0.84\pm0.18$ & $0.91$ \\
\bottomrule
\end{tabular*}
\end{table*}

\end{document}